# A Deep Mixing Solution to the Aluminum and Oxygen Isotope Puzzles in Presolar Grains

S. Palmerini [1,2],[*] O. Trippella[1,2], and M. Busso[1,2]

[1] Department of Physics and Geology, University of Perugia, Via A. Pascoli, 06125 Perugia, Italy

[2] INFN, Section of Perugia, Via A. Pascoli, 06125 Perugia, Italy

**ABSTRACT**

We present here the application of a model for a mass circulation mechanism in between the H-burning shell and the base of the convective envelope of low mass AGB stars, aimed at studying the isotopic composition of those presolar grains showing the most extreme levels of $^{18}O$ depletion and high concentration of $^{26}Mg$ from the decay of $^{26}Al$. The mixing scheme we present is based on a previously suggested magnetic-buoyancy process, already shown to account adequately for the formation of the main neutron source for slow neutron captures in AGB stars. We find that this scenario is also capable of reproducing for the first time the extreme values of the $^{17}O/^{16}O$, the $^{18}O/^{16}O$, and the $^{26}Al/^{27}Al$ isotopic ratios found in the mentioned oxide grains, including the highest amounts of $^{26}Al$ there measured.

## 1 INTRODUCTION

Chemical peculiarities involving CNO and intermediate-mass isotopes, observed in stellar photospheres (see e.g. Gilroy & Brown 1991), prove that non-convective transport mechanisms occur in stars. In particular, for low mass red giants, several authors (see e.g. Charbonnel 1994, 1995; Wasserburg et al. 1995; Eggleton et al. 2006, 2008; Karakas et al. 2010; Busso et al. 2010; Palmerini et al. 2011b,a) have shown that the first dredge-up (hereafter FDU), which determines the surface composition at the beginning of the Red Giant Branch (RGB), and the third dredge-up (hereafter TDU), which occurs after each He-burning runaway episode (thermal pulse, or TP) of the Asymptotic Giant Branch (AGB) phase, are insufficient to explain the observed abundances of several nuclei from $^3$He to $^{26}$Mg.

Of particular importance in helping us to constrain the nucleosynthesis and mixing mechanisms occurring in red giants is the record of isotopic abundances measured in presolar grains, which are tiny particles of dust, found in pristine meteorites and formed in circumstellar envelopes. Whenever we have independently some indications on their parent source (otherwise a priori unknown in terms of stellar mass and metallicity) they can give us very precise information about the physics of the stellar site in which they formed. A consistent part of the presolar O-rich solids are made up of corundum ($Al_2O_3$) grains, which are classified in four groups according to the oxygen isotopic mix they show (see Nittler et al. 1997, and their Figure 5 and Table 3 in particular).

Grains belonging to groups 1 and 2 are characterized by values of $^{17}O/^{16}O$ and $^{18}O/^{16}O$ respectively larger and smaller than the solar values. They can not be accounted for by the enrichment in $^{16}O$ typical of a core-collapse Supernova and are then believed to form in red giant stars (Zinner 2005). In particular, group 1 grains have $4.5 \cdot 10^{-4} \leq$ $^{17}O/^{16}O \leq 10^{-2}$ and $8 \cdot 10^{-4} \leq$ $^{18}O/^{16}O \leq 2.5 \cdot 10^{-3}$, while in group 2 grains the $^{17}O/^{16}O$ ratio ranges from $5 \cdot 10^{-4}$ to $1.5 \cdot 10^{-3}$ and $^{18}O/^{16}O$ is smaller than $8 \cdot 10^{-4}$ (down to a few $10^{-5}$). On the contrary, Group 3 grains show $^{17}O/^{16}O$ and $^{18}O/^{16}O$ values slightly smaller than the solar ones; this can be explained by the galactic chemical evolution, if these grains come from stars of metallicity slightly lower than solar and if $^{17}O$ and $^{18}O$ are of secondary nature. The stellar environment where group 4 grains formed is instead more difficult to be determined, because of their $^{18}O/^{16}O$ values up to ten times larger than the solar ones. However, the multi-element isotopic mix found in part of those grains suggest supernova origins (see Figure 1 in Nittler 2009, and references therein).

Most small abundances of $^{18}O$ and several large excesses of $^{26}Mg$ found in group 1 and 2 grains are not explained by pure dredge-up phenomena, and decisively contribute to the general conviction that non-convective transport mecha-nisms coupled with nucleosynthesis exist and are important (see e.g. Busso et al. 2010, and references therein).

Wasserburg et al. (1995) suggested the presence of a deep matter circulation as a possible solution to the puzzle of the oxygen isotopic mix in $Al_2O_3$ grains as well as that of the small values of the $^{12}C/^{13}C$ ratio in low mass red giant stars. These authors developed the model of a "conveyor belt" in which currents transport matter downward, from the bottom of the cool envelope to the regions where H-burning occurs, and upward, in the opposite direction, enriching the stellar surface with fresh products of the CNO cycle. This mechanism was called "Cool Bottom Process" (hereafter CBP) in antithesis with Hot Bottom Burning, which is instead typical of more massive objects, where the temperature at the base of the convective envelope is high enough to allow proton capture reactions to take place.

The $^{26}Mg$ excess found in presolar grains hints at their formation in $^{26}Al$-rich environments. Boothroyd et al. (1994) suggested their stellar progenitors to be late AGB stars of about $1.5M_\odot$, which had experienced many TDU episodes. They also noticed that the $^{26}Al/^{27}Al$ ratio shown by several grains is higher than the level allowed by TDU, which can typically provide $^{26}Al/^{27}Al$ ratios up to a maximum of few parts per mil; see e.g. Wasserburg et al. (2006) and the on-line FRUITY database described in Cristallo et al. (2011). On the contrary, since the first measurements, Al ratios in oxide grains showed values ranging up to a few $10^{-2}$ (Nittler et al. 1997). Although Izzard et al. (2007) noted that reaction rate uncertainties can play a role here, the subsequent analysis performed in Iliadis et al. (2011) and in Palmerini et al. (2011a) showed that this is not sufficient to explain the high values measured for the $^{26}Al/^{27}Al$ ratio. Moreover, a subsequent work by the LUNA (Laboratory for Underground Nuclear Astrophysics) collaboration (Straniero et al. 2013) pointed out that revisions of the rate for proton captures on $^{25}Mg$ might make even CBP (as modelled parametrically at that moment) insufficient to explain the highest ratios. Also the very uncertain p-capture destruction rate of $^{26}Al$ was recently revised. This rate is controlled by the 127 keV resonance in $^{27}Si$ (Margerin et al. 2015), which was remeasured and found to be a factor of 4 higher than the previously adopted upper limit (Pain et al. 2015). Even so, the effects of p-captures remain negligible: changing their rate by a factor of four upward changes the $^{26}Al$ production by only 1%, due to the low temperatures typical of shell H-burning in low mass stars.

In the past, Nollett et al. (2003) clarified that, because of the long half-life of $^{26}Al$ (about $7.17 \cdot 10^5$ years) its enrichment is not significantly dependent on the mixing velocity nor on the mixing rate, while it is strongly related to the maximum depth of mixing penetration. Indeed the isotope is produced by the H-burning shell via the $^{25}Mg(p,\gamma)^{26}Al$ reaction, which efficiently burns only at relatively high temperature.

Several group 1 and 2 oxide grains were found to lie in regions of the $^{17}O/^{16}O$ versus $^{18}O/^{16}O$ diagram that were considered as inaccessible to nucleosynthesis models (even including CBP) until a few years ago, when the reanalysis made by Palmerini et al. (2011a) using improved nuclear physics measurements (Adelberger et al. 2011; Iliadis et al. 2010) revealed that (i) the isotopic ratios of these grains could not be produced in massive AGB stars[1] and (ii) a match was actually possible between CBP models and the measured oxygen abundances even for the previously unreachable area.

Specifically, the curves of the extra-mixing models presented by Palmerini et al. (2011a) in their Figure 11a, pass through the region of the $^{17}O/^{16}O$ vs $^{18}O/^{16}O$ diagram occupied by several group 2 grains, where the $^{17}O/^{16}O$ values range between 0.0005 and 0.0015 and the $^{18}O/^{16}O$ ones are smaller than 0.0015; this is the region previously unreachable on the basis of older reaction rate compilations (see Figure 16 in Nollett et al. 2003). The new findings were then confirmed by Palmerini et al. (2013). Therefore, RGB stars with masses from 1 to $2M_\odot$ and nearly solar composition are now considered to be the progenitors of group 1 grains, while AGB stars with masses smaller than $1.5M_\odot$ are believed to be at the origin of the oxygen isotopic ratios measured in group 2 grains, thanks to more efficient CBP mechanisms (Busso et al. 2010).

Despite the success of parameterized extra-mixing models in explaining both the isotopic ratios of CNO elements observed in stellar photospheres (Hedrosa et al. 2013; Abia et al. 2012) and the isotopic abundances of $^{17}O$ and $^{18}O$ in presolar dust, reproducing with them values of $^{26}Al/^{27}Al \geq$ 0.01 requires to push the mixing well inside the H-burning shell, to layers so deep and hot that they would strongly affect the stellar energy balance, with relevant luminosity feedbacks. If this were the case, then the same post-process approach often adopted for deep mixing simulations would be no longer acceptable and full stellar model runs would be necessary. However, given the need of extended sets of test calculations, typical of the fine-tuning of free parameters, the amount of work required would become large and demanding, and one should demonstrate that any change thus induced in stellar evolution does not violate known properties of evolved low mass stars (hereafter LMS). Hence an explanation for the largest $^{26}Al/^{27}Al$ ratios based on mixing seemed initially hard to find. On the other hand, the possibility of obtaining solutions entirely based on revisions of the nuclear physics inputs seemed since the beginning excluded, both for oxygen and alluminum isotopic ratios.

One has to remember that, until recently, the physical mechanism at play in non-convective mixing was completely unknown. Although the existence of deep mixing was clearly established and several approaches for its modeling had been presented, no effective physical trigger had been proven to work in driving enough transport to account for the constraints of both stellar observations and presolar grain composition. Rotational effects or diffusion-like mechanisms were extensively explored (Zahn 1992; Palacios et al. 2006), but with non-conclusive results. Among diffusive processes, a popular one was for a few years the so-called "thermohaline mixing", induced by the activation of the $^3He(^3He,2p)^4He$ reaction, which locally reduces the molecular weight in H-burning regions (Eggleton et al. 2006; Charbonnel & Zahn 2007; Charbonnel & Lagarde 2010).

---

[1] Using in stellar models the most recent $^{16}O(p,\gamma)^{17}F$ rates one finds that values of $^{17}O/^{16}O$ smaller than $1.76 \cdot 10^{-3}$ are not accessible to nucleosynthesis from those stars, as demonstrated by Iliadis et al. (2008).

This is an unavoidable consequence of the stellar nuclear processes and was suggested as an explanation for anomalies observed in environments of evolved stars (Eggleton et al. 2008; Stancliffe 2010; Angelou et al. 2012). However, it was subsequently proven to be ineffective for abundance changes (Traxler et al. 2011) and in general too slow (Denissenkov & Merryfield 2011). Moreover, as far as the subject of this work is concerned, thermohaline diffusion is certainly inadequate to explain any change in $^{26}Al$ along the AGB. Indeed, when a star ascends the giant branch for the second time, its internal layers where the $^3He+^3He$ reaction prevails over the $^3He+^4He$ one (determining the local inversion of the molecular weight) are too shallow and cool to host any $^{26}Al$ production (see Palmerini et al. 2011a, and references therein). In any case, it is clear that something better than fully parameterized models should be developed if we want not only to explain what was observed so far but also to provide new predictive tools for stellar evolution. In fact, while circulation and diffusion models critically depend on two main parameters (the circulation rate, or the diffusion coefficient, and the depth), they hide many more degrees of freedom. One has for example to make guesses on what is the fractional area at the interface between the radiative and the convective regions that is occupied by the upward and downward flux of material; and also what is the trigger and on which parameters it depends. Hence, at a closer look, the number of almost free assumptions rapidly grows.

The above facts stimulated efforts aimed at modelling more effective mechanisms in scenarios also capable of reducing the number of free parameters. Among them, the circulation induced by a stellar dynamo, as discussed by Busso et al. (2007), Nordhaus et al. (2008) and Denissenkov et al. (2009) seemed to offer some promising chances. In the past stellar magnetic fields and their instabilities have been the object of detailed analysis (see e.g. Spruit 1999, 2002), more recently Nucci & Busso (2014) and then Trippella et al. (2016) showed how the thermodynamical conditions of the stellar layers below the convective envelope in low mass red giants allow for a natural expansion of magnetized zones, carrying matter from near the nuclear-burning shell to the envelope and then pushing down envelope matter to the radiative zone for mass conservation.

A necessary condition for the suggested solution to hold is that the layers of interest be represented very closely by polytropic structures of a rather large index (like e.g in the case of bubbles dominated by radiation) and the local density distributions follow a steep power-law of the radius, in the form $\rho \propto r^k$ (where $k$ is a negative number with a modulus significantly larger than unity). These conditions are not met by the Sun, so that most of the enormous work done on solar magnetic fields and their instabilities cannot be fruitfully exploited.

Establishing clear connections between stellar magnetic fields and mixing phenomena has been so far hampered by the fact that internal fields in red giants were considered as inaccessible to direct observation, with no real proof of their existence at the intensities required (i.e. $B \geq 10^5$ Gauss, see Busso et al. 2007). However very recently, asteroseismology studies made by the *Kepler observatory* suggested the presence of strong internal magnetic fields in LMS, through their effects in suppressing dipole oscillatory modes (Fuller et al. 2015). Many of the field values inferred were in the range originally suggested to produce the required mixing.

Also (but not exclusively) in the light of these suggestions we apply here the formalism developed by Nucci & Busso (2014) to the radiative region located below the convective envelope of low mass AGB stars, when the H shell is burning. Our attempt is to verify whether the buoyancy of magnetized domains can drive radiative mixing to account for the expected changes in the composition of the envelope and for the $^{17}O/^{16}O$, $^{18}O/^{16}O$, and $^{26}Al/^{27}Al$ isotopic ratios measured in oxide presolar grains, without introducing changes in the energy balance of the star (hence without the need to abandon a simple post-process approach). In Section 2 we discuss our model and its application to the specific problem presented here. Then in Section 3 we compare our predictions with the measurements in SiC grains, while our conclusions are outlined in Section 4.

## 2   A MODEL FOR EXTRA-MIXING ABOVE THE H-SHELL OF EVOLVED LMS

As mentioned, Nucci & Busso (2014) showed that the full MHD[2] equations can be simplified considerably, thus becoming analytically solvable, for the very peculiar physical conditions prevailing below the convective envelopes of low mass stars. This is so because three special situations occur in those stellar regions:

(i) the plasma density distribution can be well approximated by a power law $\rho \propto r^k$, where r is the stellar radius and $k$ is smaller than $-1$;

(ii) the magnetic Prandtl number $P_m$ (namely the ratio between the kinematic viscosity $\eta = \mu/\rho$ and the magnetic diffusivity $\nu_m$) is larger than unity in the region where the expansion of magnetized zones starts (see Spitzer 1962);

(iii) while the kinematic viscosity $\eta$ cannot be really neglected, the dynamic viscosity $\mu$ remains rather small, due to the low density.

In particular, Nucci & Busso (2014) adopted a model for a $1.5M_\odot$ star from the FRANEC code and showed (in their Figure 1) that in the radiative layers below the convective envelope the physical state was such that $k$ was close to $-3$; the pressure scaled as $r^{k-1}$ and the temperature scaled as $r^{-1}$. From the equation of state of a perfect gas one then sees that the pressure roughly follows a relation of the type:

$$P \propto \rho^{4/3}. \tag{1}$$

Such a polytrope, with its low exponent $\delta \simeq 4/3$ (polytropic index $n \simeq 3$) is similar to a bubble of radiation. The existence of a simple analytical solution of a quasi-ideal MHD, hence, can be interpreted as being due to the fact that, while the relevant layers are rather extended in radius (about 60-70% of a solar radius), they contain very little mass, so that radiation dominates and all the dissipative effects induced by the presence of matter (either neutral or electrically charged, i.e. viscosity, thermal and magnetic diffusivity, etc...) are small. This fact answers the warnings advanced on the velocity of magnetic buoyancy by Denissenkov et al. (2009), permitting a rather fast (from several cm/s to several m/s) mixing process; this remains in any case considerably slower than convective motions. We underline that $k$ is not a free parameter, but a measure of how much the structure is controlled by radiation thermodynamics, something that is dictated by the stellar model.

---

[2] Magneto-hydrodynamics.

In this work we adopt stellar models for low mass stars (in the range $1.2 - 1.5\,M_\odot$) computed in 2011 by S. Cristallo for the Palmerini et al. (2011b,a) papers (see Cristallo et al. 2011, for more details). Such models predict a structure, below a red giant convective envelope, very similar to that of the older models used by Nucci & Busso (2014). This is illustrated in Figure 1, where the specific case of a $1.5M_\odot$ model is shown. The polytropic distribution mimicking the structure of the radiative layers has again an exponent rather close to 4/3 and the discussion of Nucci & Busso (2014) applies directly. We suspect that this might be a general rule, with only small differences from one stellar code to the other, which would indicate MHD processes as promising mixing mechanisms from a broader point of view.

We notice that, for the explored mass range, the thermodynamic properties of the relevant radiative layers are rather similar to each other; this however does not apply to the convective envelopes. In particular, only the most massive model considered here ($1.5M_\odot$) undergoes some TDU episodes; lower masses have the surface abundances changed only by convection affecting H-rich layers and by deep mixing, which samples the distribution of abundances established by partial H-burning above the H-burning shell (see e.g. Figure 1C); therefore, only proton-capture nuclei change gradually their abundances. In any case, even for the upper mass limit of $1.5M_\odot$, the effects of TDU on oxygen and aluminum abundances are very small.

In our stellar mass range, while the H shell is burning, the density distribution in the radiative region below the convective envelope has the form required: $\rho \propto r^k$ and $k$ remains rather close to $-3$, down to very deep layers (Figure 1A and B). The magnetic Prandtl number $P_m$ is not much larger than 1 at the base of the convective envelope, but its value increases strongly in the layers immediately above the H-burning shell, from which buoyancy can therefore start, ensuring a quasi-free MHD state (Spitzer 1962). Finally, the magnetic diffusivity $v_m$ is always very small (see Figure 1B)[3]

After checking that the above rules are in fact satisfied, we assumed magnetic fields to remain constant in time, so that the simplest version of the analytical solutions by Nucci & Busso (2014) can be used and the radial component of the outflow velocity is given by:

$$v_r(r) = v_{r,p}\left(\frac{r}{r_p}\right)^{-(k+1)} \qquad (2)$$

and applying the simplest relation between $v_r$ and $B_\varphi$, which is set by arbitrary functions, one can safely write the magnetic field in the form:

$$B_\varphi(r,t) = B_{\varphi,p}\cos(\omega t + r_p/r)\left(\frac{r_p}{r}\right)^{-(k+1)}. \qquad (3)$$

where $-(k+1) \sim 2$, being $k \sim -3$ and $\Phi$ is a function of time, measured in units $[\mathrm{Gauss} \cdot \mathrm{cm}^{-(k+1)}]$. Here $w(t)$ is a function whose derivative in time is the previously introduced function $\Gamma(t)$ (see again Nucci & Busso 2014).

We notice here that, while $k$ is not strictly a free parameter (in the sense already specified), its value changes slightly in different evolutionary stages and for different stellar masses. This means that our models still have some "parameterizations". This depends on the fact that the solutions hold when "quasi-free MHD" conditions occur, but this forcedly does not offer a very stringent constraint in terms of the starting layer $r_p$ (hence of $k$ itself).

We use the suffix "$p$" to indicate the values of the parameters pertaining to the layer from which buoyancy (on average) starts. The choice of this notation is inherited by Wasserburg et al. (1995); Nollett et al. (2003); Palmerini et al. (2011a), who in their models indicated as $T_p$ the temperature of the deepest layers of the H-shell reached by the transported material.

The parametric models of the authors quoted above were essentially based on a fine tuning of two main free parameters: the amount of material mixed into the stellar envelope and the maximum penetration of the mixing. The first one depends in its turn both on the velocities of the upward and downward streams of matters and on the fraction of the area, at fixed radius, occupied by the "conveyor-belt". In our more physical approach $r_p$ represents the zone where MHD changes from resistive to quasi-ideal and the magnetic structures can start to be lifted in a free expansion, according to the exact solution of the MHD equations. As stated, this cannot be translated into a unique, fixed layer. As an example, choosing the zone where $P_m$ is in the range from 8 to 15 would mean choosing values of $r_p$ between $0.03-0.1R_\odot$. Using these values in a formula like Equation 2 and finding best fits, one derives values of the exponent $k$ ranging from $-3.1$ to $-3.5$, with regression coefficients always larger than 0.998. This is equivalent to saying that there is actually a region of finite thickness $\Delta r_p$ from where the magnetized structures can start their buoyancy. In each specific layer of this zone $^{26}Al$ has a different abundance, due to its strong dependence on the temperature (hence on the stellar position). In what follows, therefore, we shall consider not one but a certain number (five) of starting layers $r_{p,i}$, each one associated with its corresponding exponent $k_i$ in Equation 2. With the above approach, buoyant structures satisfying the criteria by Nucci & Busso (2014) will leave the internal zones with abundances depending on the starting point. A sketch of the varying composition of the relevant layers is shown in Figure 1C. It also turns out that this composition is not further changed in the flow. This fact is crucial for the changes in the model surface abundances which the resulting mixing can lead to. In fact, the time spent by matter in traveling toward the envelope is too short to allow for nuclear reactions. It actually takes only about 1 year to the magnetic instabilities to cross the radiative layers; and less than 3-4 months are spent at temperatures higher than $3 \cdot 10^7$ K. Hence, The physics at the base of the process, near the H-shell, controls what is obtained at the surface. We underline that, despite its simplicity, this model is a remarkable improvement over the original, continuous and parameterized "conveyor belt". In that case matter had time to "cook" along the path, in streams usually rather slow and taking several years to reach the envelope. Then the most important parameter of the model was actually the path integral of the reaction rates along the transport trajectory.

---

[3] Following the approaches by Chapman & Cowling (1969); Parker (1960); Spitzer (1962), and Nucci & Busso (2014), the dynamical viscosity and magnetic Prandtl number can be described by the formulae: $\mu = \lambda c_s \rho$ and $P_m \simeq 2.6 \cdot 10^{-5}T^4/n$, where $c_s = \sqrt{\gamma(P/\rho)}$ is the sound speed, $\gamma$ is the adiabatic exponent, $\lambda$ is the De Broglie wavelength for the proton and $n$ is the number density of the plasma.

We notice that the effects of mixing mechanisms different from MHD processes are not included in this paper. This is in particular true for rotation, despite the fact that establishing a dynamo requires that a (differential) rotation occurs. The reasons for this choice are twofold. On one side, the analysis performed by Charbonnel & Lagarde (2010) showed that rotational mixing, although important in first-ascent red giants for explaining CN values, requires long time scales, being characterized by extremely small diffusion coefficients (see e.g. Figure 9 in that paper). This makes it ineffective as compared to the typical time scale for buoyancy (about 1 year). The second reason is that our model, descending from those by Nucci & Busso (2014), a priori assumes that rotation is in fact slow, so that the stellar structure is not distorted by it. This is a necessary condition for finding an exact analytical solution. Concerning thermohaline diffusion, as we have already mentioned, it is in its turn too slow and also too shallow to affect $^{26}$Al.

In the parametric mixing scenario described by Wasserburg et al. (1995), and later by Nollett et al. (2003) and Palmerini et al. (2011a), upflowing streams of material are triggered (because of the mass conservation law) by the penetration of matter from the bottom of the stellar envelope; instead, Nucci & Busso (2014) suggest magnetized materials to be pushed up into the envelope from below (driven by magnetic buoyancy) and the downflow to occur as a consequence of mass conservation at the envelope border. Once magnetic structures cross the upper boundary of the radiative region, macroturbulence in the convective envelope rapidly destroys them and the trapped mass is released into the whole envelope. Flux tube destruction is essential to modify the stellar surface composition by the introduction of fresh products from the H-burning shell. In order for this to happen, the speed by which magnetized structures penetrate into the envelope ($v_{r,e}$) has to be low enough as compared to the *average* convective velocities in the bottom envelope layers ($v_{conv}$). On the basis of previous studies of buoyant magnetic fields at the base of convective envelopes (Rempel 2003; Fan 2008) we assumed $v_{r,e} \simeq 100$ m/s, as in Nucci & Busso (2014). This is motivated as follows. In the sub-adiabatic regions at the bottom of the envelope, an estimate of the rise velocity of flux tubes was derived by Rempel (2003). Assuming a quasi-stationary equilibrium between the heating and the stratification effects, the relation $v_{r,e} \simeq 10^{-5}/|\delta|$ m/s was derived. Here $\delta = \nabla - \nabla_{adiab}$. The lifetime of tubes in that case was found to be $\tau \simeq -1.1 \cdot 10^7 \delta$. Imposing that the flux tubes are really rapidly destroyed (e.g. $\tau = 1$ day) then yields $\delta \simeq -10^{-7}$, hence $v_{r,e} \simeq 100$ m/s. In any case, the maximum values attained by the mixing velocity are always considerably smaller than the Alfvén velocity, which is the speed of MHD waves.

We note that the values of $r_e$ always cover a substantial fraction of a solar radius. As an example, a value of $4.5 \cdot 10^{10}$ cm (about $0.65 R_\odot$) pertains to our $1.5 M_\odot$ model at mid-AGB evolution. It was about 10% larger in the models adopted by Nucci & Busso (2014). From Equation 2, applied between $r_p$ and $r_e$ at mid-AGB, assuming $v_{r,e} \simeq 10^2$ m/s $< v_{conv}$, and taking an average value for $r_p$ from the stellar model adopted (about 2 to 3 hundredths of a solar radius), the velocity $v_{r,p}$ will be between 0.15 and 0.25 m/s, depending on the case: it scales as $(r_p/r_e)^2$, whenever $k = -3$, and this ratio has values of 1 part over 21-26.

Let us now make a comparison between the mass circulation rate $\dot{M}$ we can derive with our velocity profiles obtained from MHD and that previously found in parameterized models. Assuming that the value of $\dot{M}$ due to buoyancy is constant, computing the rate of mass penetration into the envelope requires an estimate for: (i) the amount of matter transported by each tube; and (ii) the fractional area these tubes occupy, when the stellar radius is $r_e$. Extrapolating solar conditions to an evolved star, a field of a few $10^5$ Gauss may be involved in magnetized structures occupying a fraction $f_1$ of $1-2\%$ of the whole area at the base of convective envelope (Nucci & Busso 2014, and references therein). Moreover, the filling factor $f_2$ of each flux tube is of about $1/100$ of its section; (they are actually almost empty, with electrically-charged materials concentrated in thin current sheets, see e.g. Hirayama 1992). Even for these parameters an a priori value cannot be determined. We must rely on studies of the Sun and then extend these results to AGB stars; again, our approach turns out to avoid "some" free parameters, but not to be completely parameter-free (something that is extremely rare).

In the light of what has been discussed so far, the mixing rate forced by magnetic buoyancy will be:

$$\dot{M} = 4\pi \rho_e r_e^2 v_{r,e} f_1 f_2. \tag{4}$$

By applying this equation to AGB stars of $1.2 - 1.5 M_\odot$ and using the above estimates for $f_1$, $f_2$ and $v_{r,e}$, the $\dot{M}$ values obtained span the interval $(0.3 - 1.3) \cdot 10^{-6} M_\odot/yr$, which is in the range found previously by Palmerini et al. (2011a) to account for the composition of group 2 oxide grains.

The downflows of matter from the envelope, pushed down by the rising material, was analyzed in detail by Trippella et al. (2016) for the more internal layers of the He-rich zones. In fact, the formation of the $^{13}$C, the subsequent neutron release via the $^{13}$C($\alpha$,n)$^{16}$O reaction and the resulting s-process nucleosynthesis depend on the amount of mass transported and on the profile of the proton penetration in the He-rich zone from the convective border, when the H-burning shell is extinguished and TDU is active.

It must be noticed that on average the mass of the downflowing material $dM_{down}$ must be equal to the mass $dM_{up}$ transported upward by the buoyancy. Then we can write:

$$dM_{up} = 4\pi \rho_e r_e^2 f_1 f_2 v_{r,e} dt \tag{5}$$

and

$$dM_{down} = 4\pi \rho_e r_e^2 v_{r,d} dt \tag{6}$$

where $v_{r,d}$ is the downward velocity at the envelope border. Hence, the constraint that $M_{down} = M_{up}$ implies that:

$$v_{r,d} = f_1 f_2 v_{r,e} \tag{7}$$

This means that the descending material, no longer confined into magnetized structures, travels to the inner stellar layers with a speed $v_{r,d}$ four orders of magnitude smaller than the rising velocity through the envelope border. In other words, it diffuses down toward the H-shell, so that the condition of Equation 7 is sufficient and we do not need to specify new "fractional areas" occupied by the descending material.

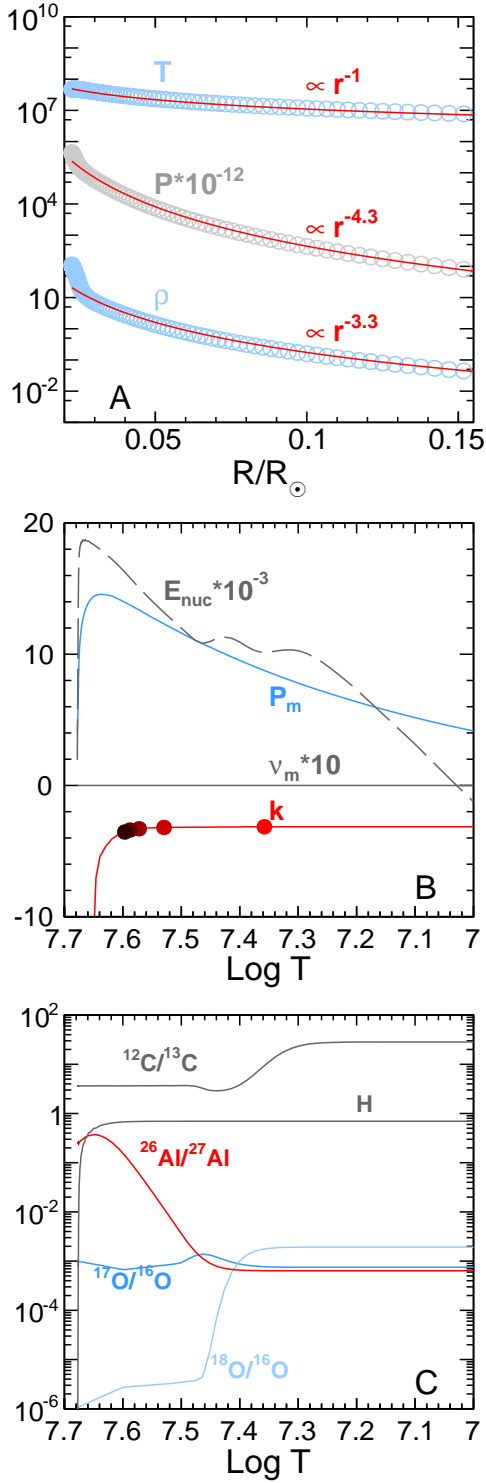

**Figure 1.** Panel A. The behavior of density, temperature and pressure as a function of the radius, in the innermost radiative zone between the H-burning shell and the base of the convective envelope (from left to right) at mid-AGB evolution of a $1.5M_\odot$ and solar metallicity model star. The curves are well represented by power laws with exponents such that a polytropic relation $P \propto \rho^\delta$, with $\delta$ close to 4/3 is maintained. Panel B. The nuclear energy generated $E_{nuc}$, the magnetic diffusivity $\nu_m$ and the magnetic Prandtl number number $P_m$ as a function of temperature. The structure is always represented well by a relation $\rho \propto r^k$, where $k$ is close to $-3$. Heavy dots represent a range of possible starting layers for magnetic buoyancy; over this range, the best-fit value of $k$ varies slightly, covering the interval from $-3.1$ to $-3.5$ (the darker the dot, the smaller the value of $k$). Panel C. The values of the isotopic ratios for carbon, oxygen and aluminum over the same temperature range of the previous panel.

We notice that if the average downflow velocity is in general much smaller than the upflow one, one can expect that mass conservation is preserved only on average, and that occasional "voids" can be created, which should be immediately compensated by fast episodes of mass downflows. In fact such a situation is exactly the one normally occurring in the Sun. Fast downstream motions compensating local mass unbalances are observed frequently (see Shimizu et al. 2008). This however has no effects on our scheme for the buoyancy of H-burning ashes, because whatever is the mechanism of matter readjustment at the envelope border, it does not imply changes in nucleosynthesis, being the local temperature too small.

## 3   COMPARISON WITH OBSERVATIONS

Figure 2 shows the comparison between the oxygen isotopic mix found in oxide grains of groups 1 and 2 and the predictions of our magnetic mixing calculations, made for the RGB and AGB phases of two representative models, that of a $1.2M_\odot$ (panel A) and that of a $1.5M_\odot$ (panel B). The two panels represent updated versions of Figure 11a in Palmerini et al. (2011a), for the chosen stellar models.

The dashed grey line shows the surface oxygen isotopic mix in stars of mass from $1M_\odot$ to $1.7M_\odot$ and solar-metallicity that ascend the red giant phase. If no extra-mixing episode occurs, these abundances are expected to remain unchanged as determined by the FDU during the whole RGB phase. Viceversa, if deep mixing takes place, it forces the isotopic ratios to cover an area below the grey curve, describing descending paths. As expected, $^{18}O$ is always depleted in this process, while $^{16}O$ remains essentially untouched, because of the relatively low temperatures in the internal layers of low mass red giants and of the low proton-capture cross section.

We adopt the solar metallicity from Lodders & Palme (2009), where $Z_\odot \simeq 0.0154$. The same nuclear physics inputs of Palmerini et al. (2013) are used, except for the rate of proton captures on $^{25}Mg$, which is upgraded, including the results by LUNA (Straniero et al. 2013).

No further free parameters are assumed in our calculations, with respect to those discussed in section 2. Indeed, the velocity profile and the mixing rate $\dot{M}$ are established by Equations 2 and 4, respectively. Consistently with the mentioned fact that buoyancy will start from a zone thick enough to host different abundances, we performed our calculations assuming flux tubes rising from layers where $k = -3.5, -3.4, -3.3, -3.2$ and $-3.1$. The various choices of $k$ foresee different processing along the RGB phase, indicated by the filled squared symbols. Hence AGB evolution starts with isotopic admixtures in the range covered by those symbols and will proceed further down, pointing towards CNO equilibrium $^{17}O/^{16}O$ ratios and complete $^{18}O$ destruction. The figure makes clear that deep mixing on the RGB can be responsible only for some, not for all the measured data. In this we confirm previous results by Busso et al. (2010).

We notice in particular that the sequence of curves from the two stellar models, during the AGB phases, cover the whole area where group 2 oxide grains lie. We get therefore a confirmation of what was suggested by Palmerini et al. (2011a, 2013); these conclusions are now based on a deep-mixing mechanism rooted in a more physical approach, and based on necessary processes of plasma physics.

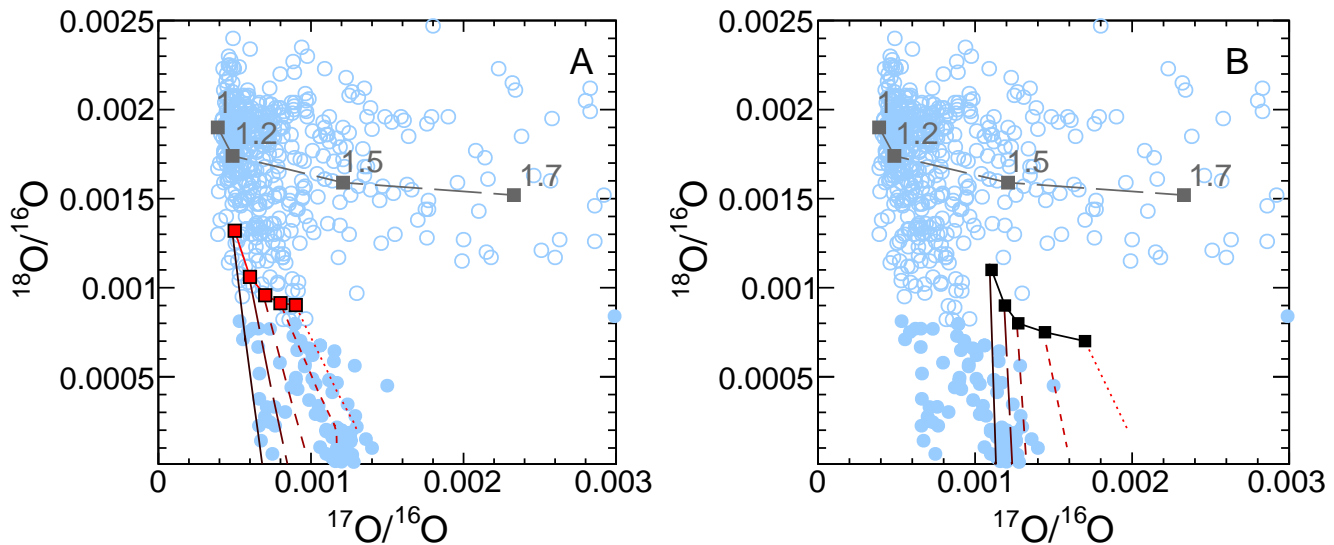

**Figure 2.** Panel A. Oxygen isotopic mix in the envelope of solar-metallicity stars at FDU (dashed grey line). Full squares indicate the composition of stars with mass from 1 to 1.7M$_\odot$, as indicated by the labels. Data points refer to measurements in presolar grains from Hynes & Gyngard (2009). We plot those of group 1 (open circles) and those of group 2 (filled circles). We do not report error bars in the figures, because they would confuse the plot, due to the large number of grains. Uncertainties are typically lower than 5-6% for all the grains of groups 1 and 2. The descending continuous lines refer to model results for extra mixing driven by magnetic buoyancy in a 1.2M$_\odot$ star during the AGB phase for different values of $k$, starting from the end of the shift introduced by deep mixing during the RGB phase (marked by the heavy squares). The smaller the value of $k$, the darker the curve, the shorter the line strokes ($k = -3.5$ solid line, $-3.4$ long dashed line, $-3.3$ regular dashed line, $-3.2$ short dashed line, and $-3.1$ point dashed line). Panel B shows the same picture, but for a 1.5M$_\odot$ stellar model.

In Figure 3, panels A, B, C, D, the evolution of the oxygen isotopic ratios is shown as a function of the $^{26}$Al/$^{27}$Al ratio for the same models discussed so far. The curves deal with the same magnetic mixing cases shown in Figure 2. In Figure 3 we see that the success of magnetically-induced mixing lays in accounting for the range of values of $^{26}$Al/$^{27}$Al up to 0.1, even adopting the already-quoted LUNA rate for $^{26}$Al/$^{27}$Al production. We recall that previous parametric models could not, instead, explain the whole range $^{26}$Al, as measured in presolar grains and were therefore excluded by Straniero et al. (2013) (see also Palmerini et al. 2011a, and in particular their Figure 11).

Moreover, observing the panels of Figures 2 and 3 one notes that also a portion of isotopic ratios of group 1 grains (open circles) is matched by the curves produced by magnetic buoyancy models during the AGB phase. Group 1 grains were suggested to form in the envelopes of RGB stars, but we see now a more continuous transition between the two cases, so many measurements are difficult to be ascribed to a specific stellar evolutionary stage. Only group 2 grains, and especially those rich in $^{26}$Al/$^{27}$Al, necessarily require AGB conditions. One has to notice here that in our nucleosynthesis network, the destruction of $^{26}$Al can occur not only through its decay, but also through n or p captures. However, n-captures in the thermal pulses concern only the most massive of our model stars, that of a 1.5 M$_\odot$, undergoing some TDU episodes. Here some $^{26}$Al saved from (n,$\gamma$) destruction can be dredged up to H-rich zones, further increasing the $^{26}$Al inventory. However, we showed that the most relevant part of the oxide grains we considered here were produced in much lower mass stars, so that n-captures are never relevant.

Finally, a mention is needed of the fact that our model for the mechanisms producing the isotopic admixture of oxygen and alluminum measured in presolar oxide grains can in fact be the same required to explain the carbon isotopic ratios and the C/O abundance ratios seen on the photospheres of evolved O-rich stars. Results from a test done in this direction are shown in Table 1.

The Table shows our predictions for the C/O and the $^{12}$C/$^{13}$C ratios in the envelope of an O-rich AGB star of 1.5M$_\odot$, at different moments in its evolution. As an exemple we have chosen an average value of $k$ ($-3.3$). The composition at the end of the RGB phase (equal to the one indicated as "Early AGB") was chosen as descending from a very moderate deep mixing process occurring during the red giant phase. (Post-RGB $^{12}C/^{13}C$ ratios ranging from about 10 to about 23 are reproduced by models with $k$ values from -3.5 to -3.1).

**Table 1.** C/O and $^{12}$C/$^{13}$C ratios predicted by our most massive model with CBP, that of a 1.5M$_\odot$ star, up to the end of the AGB phase. They are compared to results from the same stellar code with no CBP.

|  | $^{12}$C/$^{13}$C | | C/O | |
|---|---|---|---|---|
| FDU | 25 | | 0.30 | |
|  | $k = -3.3$ | no CBP | $k = -3.3$ | no CBP |
| early AGB | 19 | 25 | 0.24 | 0.30 |
| mid TP-AGB | 35 | 53 | 0.43 | 0.63 |
| end TP-AGB | 51 | 79 | 0.63 | 0.94 |

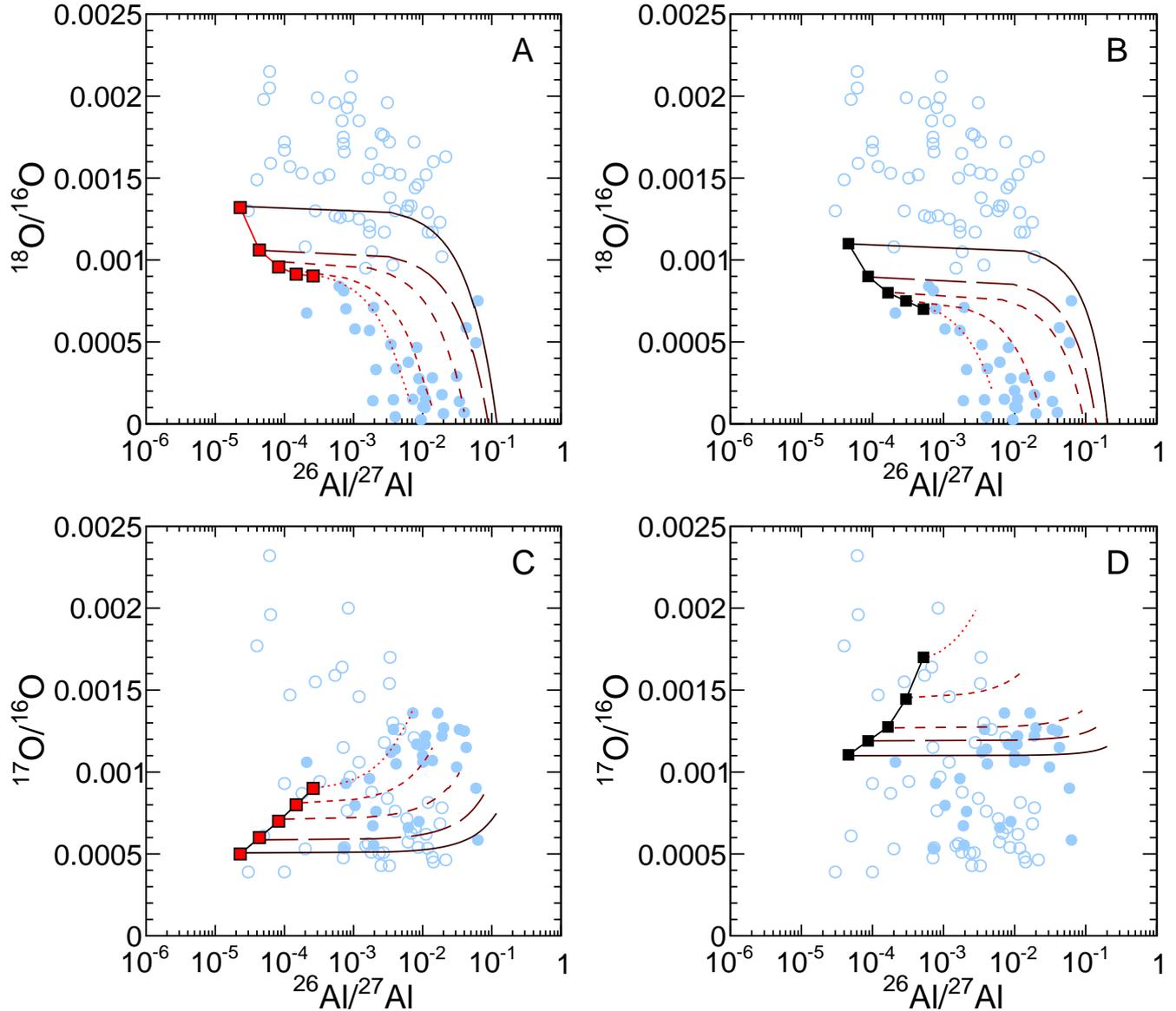

**Figure 3.** The evolution on the AGB of the oxygen isotopic ratios $^{18}O/^{16}O$ (panels A and B) and $^{17}O/^{16}O$ (panels C and D) as functions of $^{26}Al/^{27}Al$. Panels A and C refer to a $1.2M_\odot$ model, while panels B and D refer to $1.5M_\odot$ model. The heavy squares marking the starting points represent RGB compositions for different values of $k$. Model curves proceed from left to right, and are obtained for values $k = -3.5, -3.4, -3.3, -3.2$ and $-3.1$ As in Figure 2, differently dashed curves of different color refer to different values of $k$. Data grain uncertainties for the $^{26}Al/^{27}Al$ isotopic ratio are on average smaller than 25% and 10% in group 1 and group 2 grains, respectively

The subsequent evolution is obtained through the combined effects of deep mixing along the AGB, mass loss and TDU. For the sake of exemplification we show data obtained with a very slow mass loss (as used in Busso et al. 2010). With this choice the star undergoes a total of 13 thermal pulses, of which 7 with some TDU (by comparison, in the FRUITY repository, where a more efficient mass loss rate was adopted, a similar star with a similar metallicity undergoes only 11 thermal pulses, of which 5 with TDU). The data indicated in the Table as "mid TP-AGB" refer to the $4^{th}$ TDU episode ($8^{th}$ thermal pulse) and those indicted as "end TP-AGB" refer to the $7^{th}$ TDU ($11^{th}$ thermal pulse). As the figure shows, even without deep mixing, a $1.5M_\odot$ model never becomes a C-star (i.e. C/O is always below unity). The effects of deep mixing is that of limiting the growth of $^{12}C$ (which is induced by TDU), thus maintaining both the C/O and the $^{12}C/^{13}C$ ratios within the (broad and uncertain) limits observed for MS and S stars. In particular these evolved stars, with C/O ratios dispersed over a wide interval (0.3 to 0.8), have $^{12}C/^{13}C$ ratios from 15 to more than 40 (Wallerstein et al. 2011; Vanture et al. 2007), which compare well with the model values shown in Table 1.

Finally, not only carbon, oxygen and aluminum isotopic ratios are modified by the mixing. All nuclei undergoing proton captures are affected. Among the elements whose isotopic ratios might be modified one finds He, Li and N. Their detailed evolution will be the object of a forthcoming analysis, where also the effects of MHD mixing at play during the RGB phase will be investigated. We hope in this way to improve the agreement between the model predictions and the $^{17}O/^{16}O$ measurements for group 1 grains.

## 4 CONCLUSIONS

In this paper we have verified that the physics of the radiative regions above the H-burning shell is such that the buoyancy of magnetized structures can occur as a natural expansion, with a velocity profile of the form reported by Equation 2, as in previously suggested models by Nucci & Busso (2014). Our finding is that the (rather limited) constraints posed by the C/O and $^{12}C/^{13}C$ ratios typical of oxygen rich AGB stars are satisfied. This is e.g. the case of the $^{12}C/^{13}C$ isotopic ratios observed in MS- and S-stars (Wallerstein et al. 2011; Vanture et al. 2007). Concerning group 2 oxide grains, the whole areas that they occupy in the $^{17}O/^{16}O$ vs $^{18}O/^{16}O$ diagram, in the $^{17}O/^{16}O$ vs $^{26}Al/^{27}Al$ one, and in the $^{18}O/^{16}O$ vs $^{26}Al/^{27}Al$ one are covered by the model curves of our magnetic mixing. For the first time $^{17}O/^{16}O$, $^{18}O/^{16}O$, and $^{26}Al/^{27}Al$ values found in those oxide grains are reproduced together, in a physically-based model. In our approach we consider buoyancy occurring from a zone of finite thickness, where we sampled five representative positions from which magnetic structures can start to be lifted freely; these cases should correspond to a real interval of radius and mass values above the H-burning shell from which magnetized materials can detach and float, sampling different abundances in the H-burning ashes that translate into a spread of $^{26}Al/^{27}Al$ abundances induced into the envelope. Although $k$ is not strictly a free parameter, as already mentioned in section 2, we underline the cautions advanced earlier on the fact that our necessary condition of having a "quasi free" MHD process does not provide stringent constraints on $k$, so that the range of its values objectively remains a source of uncertainty in our model.

The capability of our magnetically-driven mechanism for mixing to account for the isotopic composition of group 2 oxide grains, including $^{26}Al/^{27}Al$, couples with its recent success in producing the neutron source for s-processing in adequate concentrations (Trippella et al. 2016) and offers to stellar physics a physically-based model for explaining the abundance changes occurring in evolved stars.

## ACKNOWLEDGEMENTS

This work is dedicated to the memory of our friend and teacher Gerald Joseph Wasserburg, passed away in June, 2016. Our scientific debts and human affection towards him cannot be expressed by words existing in human languages. We shall preserve forever the privilege of his friendship as one of the greatest gifts that life has granted to the three of us.

This work was partially supported by the Department of Physics and Geology of the University of Perugia, under the grant named "From Rocks to Stars", and by INFN, section of Perugia, through funds provided by the Scientific Commission n. 3 (Nuclear Physics and Astrophysics). O.T. is grateful to both these organizations for post-doc contracts. S.P. acknowledges the support of Fondazione Cassa di Risparmio di Perugia.